\documentclass[preprintnumbers,showpacs,10pt]{revtex4}

\usepackage{amsfonts,amsmath,amssymb,bm,dsfont,eucal,eufrak,slashed}
\usepackage[english]{babel}
\usepackage[latin1]{inputenc}
\usepackage[T1]{fontenc}
\usepackage{ae,aecompl}
\usepackage{graphicx,graphics,epsfig,epstopdf,subfigure,color,psfrag}
\usepackage{hyperref}

\def \de {\partial}

\def \a {\alpha}
\def \G {\Gamma}
\def \g {\gamma}

\def \s {\sigma}

\def \d {\delta}

\def \ep {\epsilon}
\def \r {\rho}

\def \m {\mu}
\def \n {\nu}

\def \f {\varphi}

\def \pr {\prime}
\def \non {\nonumber}

\def \fr {\displaystyle\frac}

\def \wt {\widetilde}
\def \ra {\rightarrow}

\def \Tr {\mbox{Tr}}
\def\laq{~\raise 0.4ex\hbox{$<$}\kern -0.8em\lower 0.62ex\hbox{$\sim$}~}%minore o circa uguale
\def\gaq{~\raise 0.4ex\hbox{$>$}\kern -0.7em\lower 0.62ex\hbox{$\sim$}~}%maggiore o circa uguale

\begin{document}

\title[]{Phenomenology Of The Holographic Soft-Wall Model Of QCD\\ With ``Reversed'' Dilaton}
\preprint{DCPT/10/158}
\preprint{IPPP/10/79}

\author{Stefano Nicotri}
\email{stefano.nicotri@durham.ac.uk}
\affiliation{Institute for Particle Physics Phenomenology, Department of Physics, Durham University, Durham DH1 3LE, United Kingdom}

\begin{abstract}
We study the phenomenology of the recently proposed modified version of the Soft-Wall model of holographic QCD with a negative dilaton profile.\\
We investigate vector and scalar mesons, evaluating mass spectra, decay constants and two-point correlation functions.
We then study chiral symmetry breaking and compute the coupling of a scalar meson with two pseudoscalars.
Finally, we compare the results with the ones found in the literature in the positive-dilaton case. 
\end{abstract}


\maketitle

%\section{Introduction}

The formulation of the AdS/CFT correspondence conjecture \cite{Maldacena:1997re} has recently generated many attempts to approach the theory of strong interactions ``from the gravity side'' \cite{gravityside}.
In particular, a class of bottom-up models has been proposed \cite{erlich,karch,Andreev:2006vy}, inspired by noncritical string theory descriptions, in which a natively non-supersymmetric framework is set up in a five-dimensional $AdS$-like space, together with a mechanism of breaking conformal symmetry, in order to catch some desired degrees of freedom of QCD at strong coupling. 
We focus on the Soft-Wall model \cite{karch,Andreev:2006vy}, which is constructed in a non-dynamical $AdS_5$ space, whose line element is, in Poincar\'e coordinates,
\begin{equation}
ds^2=\frac{R^2}{z^2}\,\left(dt^2-dx^i dx_i-dz^2\right)~~~~~z\geqslant 0~~~~~i=1,2,3\,\,.
\end{equation}
QCD is defined on the boundary $z=0$ of $AdS_5$ and the fifth ``holographic'' coordinate $z$ represents an inverse renormalization scale \cite{gravityside,erlich}.
In this model the conformal symmetry (proper of $AdS$ spaces) is broken by inserting an external non-dynamical dilatonlike profile $e^{-\phi(z)}$ in the action.
In order to obtain Regge trajectories for hadron states, $\phi(z)$ must contain powers of $z$ not higher than $z^2$ \cite{karch}, so the simplest choice is  $\phi(z)=\pm c^2 z^2$ (the parameter $c\sim\Lambda_{\mbox{\tiny QCD}}$ in the exponent is the only mass scale of the model). 
We call the correspondent models SW$^\pm$, respectively.
After the seminal papers \cite{karch,Andreev:2006vy} the SW$^+$ has been investigated in many aspects in many subsequent works.
Topics that have been studied include hadron spectra, decay constants and strong couplings \cite{BoschiFilho:2002vd,Colangelo:2008us,Colangelo:2007pt,Colangelo:2007if,vega}, form factors \cite{grigoryan}, deconfinement \cite{braga,herzog}, finite temperature effects \cite{Colangelo:2009ra,Miranda:2009uw,Fujita:2009wc,Grigoryan:2010pj}, condensates \cite{Colangelo:2007if,Forkel:2007ru}, deep inelastic scattering and structure functions \cite{DIS}.
The SW$^-$ firstly appeared in \cite{Andreev:2006ct}, to extract the static interquark potential through the investigation of the properties of an Euclidean rectangular Wilson loop.
However, in \cite{Andreev:2006ct} the negative sign of the dilaton was interpreted as an effect of the Wick rotation from the Minkowskian to the Euclidean signature.
Recently, the possibility to use the negative background in the Minkowski space for the holographic Soft-Wall model of QCD has been proposed in \cite{deTeramond:2009xk,Zuo:2009dz}.
The main motivation adducted to justify this proposal is that the positive dilaton background does not verify the sufficient conditions for the geometry found in \cite{Sonnenschein:2000qm} to produce a confining static potential.
On the contrary, switching the sign of $\phi(z)$ generates linear confinement, as shown in \cite{Andreev:2006ct} for zero temperature and chemical potential, and for $T>0$ and $\mu>0$ case in \cite{Colangelo:2010pe}.

%\section{Vector mesons, chiral symmetry breaking and scalar mesons}\label{sec:vecmes}

In \cite{karch}, the possibility of considering a negative dilaton profile had been discarded because it would have brought to a massless ground state for $\r$ mesons, which is phenomenologically not acceptable.
This is a very important point, since the main parameter of the model, the scale $c$, is usually fixed from the mass of the $\r^0$, and we will discuss it in the following.
The effective action introduced to study vector mesons and chiral symmetry breaking is \cite{karch}
\begin{equation}\label{chsbaction}
S_{eff}=\fr{1}{k}\int
d^5x\,\sqrt{g}\,e^{-\phi}\,\Tr\left\{|DX|^2+3X^2+\fr{1}{2g_5^2}\left(F_V^2+F_A^2\right)\right\}\,\,.
\end{equation}
$F_{V}^{MN}=\de^M V^N-\de^N V^M-i[V^M,V^N]-i[A^M,A^N]$, $F_{A}^{MN} =\de^M A^N-\de^N A^M-i[V^M,A^N]-i[A^M,V^N]$ are the gauge connections of the vector and axial fields $V^M=(L^M+R^M)/2$ and $A^M= (L^M-R^M)/2$, where $L_M$ and $R_M$ are dual to the four dimensional conserved currents $j_{L/R}^\m=\bar q_{L/R}\g^\m T^a q_{L/R}$.
The covariant derivative is $D_MX=\de_M X-i [V_M,X] -i \{A_M,X\}$.
The global symmetry $U(n_F)_L\times U(n_F)_R$ of QCD becomes a gauge symmetry in the bulk, with generators $T^a$'s.
$X(x,z)=(X_0(z)+S(x,z))e^{2i\pi(x,z)}$ is a scalar field dual to the $\bar qq$ operator.
$X_0$ depends only on $z$ and represents the expectation value $\langle\bar qq\rangle$, responsible for chiral symmetry breaking (to $SU(n_f)_V$); $S(x,z)=S^A(x,z)T^A=S_1(x,z)T^0+S_8^a(x,z)T^a$ is dual to the operator ${\CMcal O}_S^A=\bar qT^Aq$ and contains singlet component $S_1$ and octet components $S_8^a$ (we consider the $n_F=3$ case) \cite{Colangelo:2008us}; the phase $\pi(x,z)$ is the chiral pseudoscalar field.
The field dual to the vector current $J_V^\m=\bar q \g^\m q$, which represents the vector mesons, is $V_M (x,z)$.
Defining $B(z)=\left(\phi+\log(z/R)\right)/2$, after transforming $\psi(q^2,z)=e^{-B(z)}V(q^2,z)$, where $V(q^2,z)$ is a generic component of $V_\m$, we obtain a Schr\"odinger-like equation of motion for the rotated field $\psi(q^2,z)$ in Fourier space (in the axial gauge $V_z=0$, where the field $V_\mu$ is transverse: $\de_\m V^\m=0$):
\begin{equation}\label{schreqvec}
-\psi^{\pr\pr}+\left[c^4 z^2+\fr{3}{4z^2}\right]\psi=q^2 \psi\,\,,
\end{equation}
independently of the choice of the sign of $\phi(z)$ (the prime denotes a derivative with respect to $z$). 
This means that equation \eqref{schreqvec} must admit the same set of eigenvalues and eigenfunctions in both the SW$^\pm$.
The vector meson spectrum is \cite{karch}
\begin{equation}\label{vecmespectrum}
m_n^2=4c^2 (n+1)\,\,,~~~~~n=0,1,2\ldots\,\,.
\end{equation}
This shows that no massless ground state appears when one reverses the sign of the dilaton profile.
One can also check that the solution of equation \eqref{schreqvec} with $q^2=0$, which would correspond to a massless hadron, is not square integrable.
From the spectrum \eqref{vecmespectrum} we fix the value of the mass scale of the model, $c=m_\r /2$ for both SW$^\pm$.

This point can be further investigate by analyzing the two-point correlation function of the vector current $J_V^\m=\bar q \g^\m q$.
To do this, we need to evaluate the bulk-to-boundary propagator of the vector field, which is a regular (i.e. free of IR essential singularities) solution of \eqref{schreqvec} with boundary condition $V(q^2,0)=1$; we find $V(q^2,z)=\G(1-q^2 /4c^2)\,e^{-c^2 z^2}U(-q^2 /4c^2,0,c^2 z^2)$, where $U$ is the Tricomi confluent hypergeometric function. 
We can holographically evaluate the two-point correlator:
\begin{eqnarray}\label{Pivec}
\Pi(q^2) & \propto & \fr{e^{c^2 z^2}}{q^2 z}V(q^2,z)\de_z V(q^2,z)\biggl|_{z\to0}\\
& \propto & \fr{1}{2q^2}\left[-4c^2-q^2\left(\g+H(-q^2/4c^2)+\log(c^2/ \n^2)\right)\right]\non
\end{eqnarray}
with $\n$ a renormalization scale and $H(x)$ the harmonic number function.
The factor $-2c^2/q^2$ does not really give rise to a physical pole, since it can be subtracted as a contact term, being the function in the square brackets defined apart from an additive constant.
After this subtraction, we find that the position of the poles of $\Pi(q^2)$ coincide exactly with the mass spectrum \eqref{vecmespectrum} evaluated looking for normalizable states, as it should be. 

At this point, once one of the main reasons for the rejection of this model has been ruled out, we proceed in investigating chiral symmetry breaking and scalar mesons.
Chiral symmetry breaking was one of the first topic studied within the Soft-Wall model \cite{karch}.
Later, scalar mesons have been added to complete the picture \cite{Colangelo:2008us}. 
The equation of motion for $v(z)=2X_0(z)$ in the SW$^-$, derived from \eqref{chsbaction}, is 
\begin{equation}
v^{\pr\pr}+\left(2c^2 z-\fr{3}{z}\right)v^\pr +\fr{3}{z^2}\,v=0
\end{equation}
and its solution is $v(z)=A\,v_1(z)+B\,v_2(z)$, with $v_1(z)=e^{-c^2z^2}\,cz\,U\left(-1/2,0,c^2z^2\right)$ and $v_2(z)=e^{-c^2z^2}\,cz\,L\left(1/2,-1,c^2z^2\right)$. 
Then, one would fix $A=\sqrt{\pi}\,m_q/(Rc)$ and $B-A(1+\g_E-\ln4)/2\sqrt{\pi}=-\langle\bar qq\rangle/3=-\s/(Rc)^3$, with $m_q=2.29$~MeV and $\s=(327~\mbox{MeV})^3$ in order to describe both explicit and spontaneous symmetry breaking, as in the Hard-Wall model \cite{erlich}.
Both solutions are acceptable because they do not contain essential singularities.
We remind that in the SW$^+$,the solution for $v(z)$ proportional to $z^3$ must be discarded, since it has an infrared essential singularity; this provokes the appearance of a proportionality law between $m_q$ and $\s$, absent in QCD.
As been suggested in \cite{Zuo:2009dz}, changing the sign of the dilaton $\phi(z)$ (i.e. considering the SW$^-$) could be a way to solve this problem, absorbing the essential singularity into the background. 
Nevertheless, we observe that the problem is not really solved, since, even if both solutions are regular in the bulk, the on-shell action becomes divergent, because of the piece not proportional to $e^{-c^2 z^2}$ at $z\ra\infty$ in $v_2$.

%Hence, the singularity in the action has been moved from the field to the background, unless one invokes some infrared regularization.
We now proceed in investigating the scalar meson sector.
The bulk-to-boundary propagator of the scalar field $\wt S(q^2,z)$ is solution of the equation
\begin{equation}\label{eqscal}
\wt S^{\pr\pr}+\left(2c^2 z+\fr{3}{z}\right)\wt S^\pr +\left(q^2+\fr{3}{z^2}\right)\wt S=0
\end{equation}
with boundary condition $\lim_{z\ra0}(R/z)\,\wt S(q^2,z)=1$, and giving an IR finite on-shell action: $\wt S(q^2,z)=\G(1/2-q^2/4c^2)\,(z/R)\,e^{-c^2z^2}\,U\left(-1/2-q^2/4c^2,0,c^2z^2\right)$.
We can then get the two-point correlator
\begin{eqnarray}\label{piadscalmes}
\Pi_{\mbox{\footnotesize AdS}}^{AB}(q^2) & = & \d^{AB}\fr{R^3}{k}\,\wt S(q^2,z)\,\fr{e^{-\phi}}{z^3}\,\de_z\wt S(q^2,z)\biggl|_{z\ra0}\\
& = & \d^{AB}\,\fr{R}{2k}\,\biggl\{\left(q^2+2c^2\right)\left[1+\g_E-2\ln\left(c^2\ep^2\right)-2\psi\left(1/2-\fr{q^2}{4c^2}\right)\right]-6c^2\biggr\}\non
\end{eqnarray}
and, comparing its expansion for $-q^2\ra\infty$ with the known QCD one \cite{reinders} we can fix $R/k=N_c/16\pi^2$, neglecting non-perturbative (e.g. instanton-like) contributions.
The two-point function has a set of poles, whose positions give hadron masses, with corresponding residues
\begin{equation}\label{reggetraj}
m_n^2=2c^2\left(2n+1\right)\,\,,~~~~~~~~~~~F_n^2=\fr{N_c}{\pi^2}\,c^4(n+1)\,\,.
\end{equation}
The square roots of the residues, $F_n$, represent the hadron decay constants.
The states are organized in Regge trajectories \eqref{reggetraj}, with the same slope as scalar glueballs or vector mesons \cite{karch}.
%The ground state has a mass $m^2_{S_0}=549$~MeV.
One effect of reversing the sign of the dilaton field is to decrease the mass of the ground state, as already pointed out in \cite{Zuo:2009dz}, to $m^2_{S_0}=549$~MeV, while the residues are left unchanged with respect to the SW$^+$.
Given the above value of the mass, the ground state could be interpreted as the isoscalar $f_0(600)$ (or $\s$) meson, even if the precise position of the corresponding pole is difficult to establish experimentally because of its large width, and because it cannot be modeled by a naive Breit-Wigner resonance \cite{pdg}.
The decay constant of the ground state, obtained from \eqref{reggetraj}, is $F_0=0.08$~GeV$^2$.
The value of the four dimensional gluon condensate is extracted in the limit $m_q=0$, $\langle\a_s/\pi\,G^2\rangle\,=\,2c^4/\pi^2\,\simeq0.004~\mbox{GeV}^4$ and is left unchanged by switching the sign of the dilaton profile \cite{Colangelo:2008us}.
An important improvement brought by the SW$^-$ concerns the dimension six operator.
Here, in fact, using the factorization approximation in QCD, one can match the corresponding ${\CMcal O}\left(1/q^4\right)$ terms in \eqref{piadscalmes} and \cite{reinders}, obtaining, in the limit $m_q\ra0$, $\sqrt{\pi\a_s}\,\langle\bar qq\rangle\simeq(154~\mbox{MeV})^3$.
This result was not present in the SW$^+$, in which the two terms cannot be matched since they have an opposite sign.

Another drawback of the SW$^+$ concerns the coupling of the scalar mesons with two pseudoscalars, which turns out to be too small with respect to the one known from phenomenology.
Now we evaluate such a coupling in the SW$^-$. 
The field involved in the holographic description of the pseudoscalar mesons are the chiral field $\pi$ and the longitudinal part $\f$ of the axial field $A_\m$, defined by $A^\m=A_\perp^\m+\de^\m\f$ (in the axial gauge $A_z=V_z=0$).
Defining the pseudoscalar field $\psi^a=\f^a-\pi^a$, the part of the action involving one scalar and two pseudoscalar mesons is, for the octet contribution,
\begin{equation}
S^{(SPP)}_{eff}=-\fr{R^3}{k}\,d^{abc}\int
d^5x\,\fr{1}{z^3}\,e^{-\phi(z)}\,v(z)\,S_8^a\,\eta^{MN}(\de_M\psi^b)(\de_N\psi^c)\,\,.
\end{equation}
With the identification \cite{Colangelo:2008us} $\wt \psi_P^a(q,z)=(1/q^2)\,{\CMcal A}(0,z)(-iq^{\m}\tilde{A}^a_{{\|_{\,0}}\,\m}(q))$, where ${\CMcal A}(q^2,z)$ is the Fourier transform of the bulk-to-boundary propagator of the axial field, $\wt A_{\perp\m}^a(q^2,z)={\CMcal A}(q^2,z)\wt A^a_{\perp0\m}(q^2)$, for the $n$-th scalar radial excitation, the explicit form for the coupling with two pions is
\begin{equation}\label{gspp}
g_{S_n PP} = \fr{8R^2 c^4}{k f_\pi^2 F_n}\,\int_0^\infty dz\,v(z)\,L_n^1 (c^2 z^2)\left[\left(\de_z{\CMcal A}(0,z)\right)^2+\fr{m_n^2}{2}{\CMcal A}(0,z)^2\right]
\end{equation}
with $f_\pi$ the pion decay constant and $F_n$ given by \eqref{reggetraj}.
Using a numerical solution for ${\CMcal A}(0,z)$ and the value $f_\pi\simeq88~\mbox{MeV}$ (result of this model), for $n=0$ we find $g_{S_0 PP}\simeq345~\mbox{MeV}$. 
If we assume that the $S_0$ eigenfunction describes the $\sigma(600)$ meson, we can evaluate the corresponding width, since the $\s\to\pi\pi$ decay is dominant with repect to the only other seen one, $\s\to\g\g$.
We can then assume $\G_\s\simeq\G(\s\to\pi^+\pi^-)+\G(\s\to\pi^0\pi^0)$.
This is given in terms of $g_{S_0 PP}$, $m_\pi$ and $m_{S_0}$ as
\begin{equation}
\G_\s =\frac{3}{2}\,\G(\s\to\pi^+\pi^-)=\fr{3\,g_{S_0 PP}^2}{32\,\pi\,m_{S_0}^3}\,\sqrt{m_{S_0}^4-4m_\pi^4}\simeq6.4~\mbox{MeV}
\end{equation}
having used a pion mass value $m_\pi=144$~MeV obtained from the Gell-Mann--Oakes--Renner relation $f_\pi^2 m_\pi^2=2m_q \s$. The result is very small, at odds with the total width of the $\s(600)$ meson, which is of ${\CMcal O}(600-1000~\mbox{MeV})$ \cite{pdg}. 
Then, even if the SW$^-$ allows to solve the known drawback of the SW$^+$ of having independent description of spontaneous and explicit chiral symmetry breaking, making possible to consider the two independent solutions for $v(z)$ and to introduce separately the quark mass and the chiral condensate, the coupling of scalar mesons to two pions turns out to be too small than the one expected from known phenomenology.
One could justify this issue stating that the ground state does not represent the $\sigma$ meson (which in fact cannot be easily described as a pure $q\bar q$ meson \cite{sigmadiscussion}), but it is anyway difficult to explain a lowest lying $q\bar q$ state having such a small mass.

We conclude observing that, though in this respect it does not provide a fully satisfactory description of the light meson sector, the SW$^-$ certainly deserves to be studied more deeply and improved, since it is able to reproduce some key features of QCD that are absent in the SW$^+$.

\begin{acknowledgments}
  I would like to join the Organizers of the workshop in warmly remembering Beppe Nardulli, as a man of irony, science and hard work.\\
  This work was supported in part  by the EU contract No. MRTN-CT-2006-035482, ``FLAVIAnet". 
\end{acknowledgments}

\end{document}